\documentclass[aps,amsmath,amssymb,reprint,prb,showpacs]{revtex4-1}
\usepackage{xcolor}
\usepackage{graphicx}
\usepackage[dvips]{hyperref}
\usepackage{breakurl}
\usepackage{bm}
\begin{document}

\title{{
Multiplet effects in orbital and spin ordering phenomena: 
A hybridization-expansion quantum impurity solver study}}

\author{Andreas Flesch}
\affiliation{Institute for Advanced Simulation, Forschungszentrum J\"ulich, 52425 J\"ulich, Germany}
\author{Evgeny Gorelov}
\affiliation{Institute for Advanced Simulation, Forschungszentrum J\"ulich, 52425 J\"ulich, Germany}
\author{Erik Koch}
\affiliation{German Research School for Simulation Science, J\"ulich, Germany}
\affiliation{JARA High-Performance Computing}
\author{Eva Pavarini}
\affiliation{Institute for Advanced Simulation, Forschungszentrum J\"ulich, 52425 J\"ulich, Germany}
\affiliation{JARA High-Performance Computing}

\begin{abstract}
Orbital and spin ordering phenomena in strongly correlated systems are studied
using the local-density approximation + dynamical mean-field theory approach. 
Typically, however, such simulations are restricted to simplified models (density-density Coulomb interactions, high symmetry couplings and few-band models). 
In this work we implement an efficient general hybridization-expansion continuous-time quantum Monte Carlo impurity solver (Krylov approach) which allows us to investigate orbital and spin ordering in a more realistic setting, including interactions that are often neglected (e.g., spin-flip and pair-hopping terms), enlarged basis sets (full $d$ versus $e_g$), low-symmetry distortions, and reaching the very low-temperature (experimental) regime.
We use this solver to study ordering phenomena in a selection of exemplary low-symmetry 
transition-metal oxides: LaMnO$_3$ and rare-earth manganites as well as the perovskites CaVO$_3$ and YTiO$_3$. We find that, in all considered cases, the minus sign problem mostly appears when off-diagonal crystal-field terms are present
(and is strongly suppressed in the basis of crystal-field states), while off-diagonal terms of the hybridization function matrix are not as critical.
We show that spin-flip and pair hopping terms do not affect the Kugel-Khomskii orbital-order melting transition in rare-earth manganites, or the suppression of orbital fluctuations driven by crystal field and Coulomb repulsion. For the Mott insulator YTiO$_3$ we find a ferromagnetic transition temperature
$T_{\rm C}\sim 50$~K, in remarkably good agreement with experiments.
For LaMnO$_3$ we show that the classical $t_{2g}$-spin approximation, commonly adopted for studying manganites, yields indeed an occupied $e_g$ orbital in very good agreement with that obtained for the full $d$ 5-orbital Hubbard model, while the spin-spin $e_g$-$t_{2g}$ correlation function calculated from the full $d$ model is $\sim 0.74$, very close to the value expected for aligned $e_g$ and $t_{2g}$ spins; the $e_g$ spectral function matrix is also well reproduced. Finally, we show that the  $t_{2g}$ screening  reduces the $e_g$-$e_g$ Coulomb repulsion by about $10\%$.
\end{abstract}
\maketitle

\section{\label{sec:introduction}Introduction}
Orbital and magnetic ordering phenomena play a crucial role in the physics of strongly correlated transition-metal oxides. Their onset depends on symmetry, lattice distortions, super-exchange interaction and the form of the Coulomb tensor.  The realistic description of ordering phenomena requires the ability of disentangling the effects of all these interactions.
In recent years, the local-density approximation+dynamical mean-field theory approach\cite{lda+dmft,book2011,evabook2011} (LDA+DMFT), which combines {\em ab-initio} techniques based
on density functional theory in the local-density approximation (LDA),
and the dynamical mean-field theory \cite{Georges1996} (DMFT),  has lead to important progress
in understanding such ordering phenomena.  
It has been shown that many-body super-exchange only weakly affects the onset of the 
orbital-order to disorder transition in rare-earth manganites,\cite{Pavarini2010} while, in the presence of strong Coulomb repulsion, a small
crystal-field is sufficient to strongly suppress orbital fluctuations and stabilize orbital order.\cite{Pavarini2004,Pavarini2008,Pavarini2010}
However, the effects of subtle Coulomb  interactions, such as spin-flip and pair-hopping terms or of quantum fluctuations, e.g., charge fluctuations between half-filled $t_{2g}$ and $e_g$ states in manganites or spin fluctuations, are not yet fully understood,
while the origin of very low-temperature magnetism in multi-orbital materials remains little investigated in a realistic context.
The hybridization-expansion continuous-time quantum Monte Carlo (CT-HYB) technique\cite{Werner2006,Werner2006a,Haule2007,Lauchli2009,Gull2011,bookwerner2011} appears to date the most promising DMFT quantum impurity solver to study real materials at experimental temperatures,
although most calculations so far have been limited to high-symmetry cases or systems for which the hybridization function is diagonal (or almost diagonal) in orbital space.\cite{Werner2006,Werner2006a,Haule2007,Lauchli2009,Surer}

In the present work we study the effects of commonly adopted approximations
on the origin of orbital and magnetic order in some exemplary low-symmetry
transition-metal oxides.
To do this, we use an efficient general implementation of the CT-HYB quantum Monte Carlo (QMC) LDA+DMFT solver for systems of arbitrary point symmetry and arbitrary local Coulomb interaction.  
In our implementation we combine a general Krylov\cite{Lauchli2009} scheme,
which we use for the low-symmetry cases, with the very fast {segment} 
implementation\cite{Werner2006}  which can be used when the local Hamiltonian does not mix flavors
(i.e., spin-orbital degrees of freedom).  
In addition, we use symmetries\cite{Haule2007,Gull2011} to minimize the computational time. 
We present results for the orbital melting transition in the rare-earth manganites 
$R$MnO$_3$,  orbital fluctuations in the 3$d^1$ perovskites CaVO$_3$ and YTiO$_3$, and ferromagnetism in the Mott insulator YTiO$_3$. Finally, we investigate the regime of validity of the $t_{2g}$ classical spin approximation often adopted to describe LaMnO$_3$ and more general manganites. 

The paper is organized as follows. In Section II we briefly discuss the approach in the context of the LDA+DMFT method. In Section III we present applications to rare-earth manganites,  vanadates, and titanates. 
We show that spin-flip and pair-hopping terms do not affect the Kugel-Khomskii orbital-order transition and weakly affect orbital fluctuations in $3d^1$ perovskites. We calculate the ferromagnetic transition temperature for the Mott insulator 
YTiO$_3$ and find excellent agreement with experiments, showing that 
orbital order is indeed compatible with ferromagnetism in this material, 
contrarily to early hypothesis.\cite{Ulrich}
For LaMnO$_3$ we show that the $e_g$ two-band Hubbard model commonly used to study the system,
in which the $t_{2g}$ electrons are treated as disordered classical spins interacting
with the $e_g$ spins via the Coulomb interaction, yields results in very good agreement with the full five-orbital $3d$ Hubbard model. 
Remarkably, the agreement is not only excellent for the occupied state in the orbitally ordered phase, but also very good for the orbital resolved
$e_g$ spectral function matrix.
Finally, in the Appendix we describe the details of our implementation of the general CT-HYB solver.

\section{Model and Method}
The most general multi-band Hubbard model for transition-metal oxides is given by
\begin{eqnarray}
\label{eq:model}
{H} =&-&\sum_{i\ne i^\prime}  \sum_{\sigma,\sigma^\prime} \sum_{m, m'} t^{ii'}_{m\sigma m'\sigma'} c^{\dagger}_{im\sigma} c^{\phantom{\dagger}}_{i'm'\sigma'} \\
\nonumber
&+&\sum_{i}  \sum_{\sigma,\sigma^\prime}  \sum_{m, m'} \varepsilon_{m\sigma m'\sigma^\prime} c^{\dagger}_{im\sigma} c^{\phantom{\dagger}}_{im'\sigma'} \\
\nonumber
&+&\frac{1}{2}\sum_i\sum_{\sigma,\sigma^\prime} \sum_{m m'\tilde{m} \tilde{m}^\prime} \!\! U_{m m' \tilde{m} \tilde{m}'} c^{\dagger}_{im\sigma} c^{\dagger}_{im'\sigma'} c^{\phantom{\dagger}}_{i\tilde{m}'\sigma'} c^{\phantom{\dagger}}_{i\tilde{m}\sigma} 
\end{eqnarray}
Here $c^{\dagger}_{im\sigma}$ ($c^{\phantom{\dagger}}_{im\sigma}$) creates (annihilates) an electron with spin $\sigma$ in orbital $m$ on lattice site $i$; $t^{ii'}_{m\sigma m'\sigma'}$ are the hopping integrals and $\varepsilon_{m\sigma m'\sigma^\prime}$  the elements of the crystal-field matrix,  obtained from LDA calculations by constructing a localized Wannier-function basis.
\cite{Pavarini2004,Pavarini2005} $U_{mm'\tilde{m}\tilde{m}'}$ are the screened Coulomb matrix elements,
typically expressed in terms of the three Slater integrals $F_0$, $F_2$ and $F_4$, with $U_{\rm avg}=F_0$ (direct Coulomb interaction)  and $J_{\rm avg}=\frac{1}{14}(F_2+F_4)$ (exchange Coulomb interaction).
In the following we find it more useful to use as parameters\cite{evabook2011,cobaltates} the diagonal element of the Coulomb matrix, $U_0=F_0+\frac{8}{5} \tilde{J}$, the Kanamori exchange parameter $\tilde{J}=\frac{5}{7}J_{\rm avg}$ and the Coulomb anisotropy
$\delta {\tilde J}={\tilde J} (\frac{1}{5}-\frac{1}{9}\frac{F_4}{F_2}) / (1+\frac{F_4}{F_2})$. 
The exchange couplings for $e_g$ and $t_{2g}$ only are then $J_{eg}=\tilde{J}+3\delta{\tilde{J}}$
and  $J_{t_{2g}}=\tilde{J}+\delta{\tilde{J}}$.
We solve the model (\ref{eq:model}) with DMFT using the CT-HYB QMC approach as quantum impurity solver.\cite{Werner2006,Werner2006a,Haule2007} 
Our implementation of the CT-HYB QMC solver is discussed in the Appendix.
It works efficiently for systems of arbitrary space-group symmetry, i.e., with both a hybridization-function matrix and self-energy matrix in the full spin-orbital space. 
We optimize our code for modern massively parallel architectures
and exploit symmetries to minimize the computational time.
We use two approaches to calculate the trace which enters in the numerical evaluation
of the Green function: the  the segment approach\cite{Werner2006}
and the Krylov method.\cite{Lauchli2009} 
The segment approach is very fast but can only be used if the local Hamiltonian does not mix flavors
(spin-orbital degrees of freedom).
The Krylov procedure is instead general and scales linearly with the inverse temperature,
becoming therefore particularly efficient in the low-temperature regime.\cite{Gull2011,bookwerner2011}
Far from phase transitions, we further enhance the efficiency by {adaptively} truncating
the local trace in the Green function. \cite{Gull2011,Lauchli2009}
Further details on our code are given in the Appendix.
Our efficient implementation allows us to include in the model Hamiltonian (\ref{eq:model})
typically neglected interactions, such as spin-flip and pair-hopping terms or  spin-orbit coupling, to study models with larger number of orbitals (e.g. with the complete 5-orbital $d$ shell) and reach very low temperatures, as essential to study magnetic transitions.
In the following, we use our code to systematically compare different models and test typically adopted approximations
on the orbital and magnetic order of a selection of exemplary materials.

\section{Results}
\subsection{Orbital-order melting in rare-earth manganites}
The origin of the orbital-order melting transition \cite{oomelting} in the rare-earth manganites
$R$MnO$_3$ with the $t_{2g}^3e_g^1$ nominal electronic configuration has been debated since long. 
Recently,\cite{Pavarini2010,Flesch2012} we have shown that the many-body super-exchange interaction plays a small role in determining the orbital-order melting temperature $T_{\rm OO}$ as well as its trends
with decreasing radius of the rare-earth ions.
However, spin-flip and pair-hopping terms, neglected in previous calculations,
restore the full degeneracy\cite{erik,d2,gorelov} of the $S=1$ multiplet, and
could enhance the strength of super-exchange, or even modify the occupied orbital.\cite{Khomskii}
Furthermore previous calculations, as most many-body studies of rare-earth manganites, rely on the classical spins approximation for $t_{2g}$ orbitals.\cite{Ahn2000} In such approximation the effects of
the $t_{2g}$ spins ($S_{t_{2g}}=3/2$) on the $e_g$ states is described through a local magnetic field due to the $e_g$-$t_{2g}$ Coulomb exchange interaction and a band-width renormalization factor arising from the spatial disorder in the orientation of the $t_{2g}$ spins. However, charge fluctuations between $t_{2g}$ and $e_g$ states or $t_{2g}$ multiplet fluctuations,
not accounted for in such a model, could affect the orbital-order %
and the occupied orbital. 
In this section we use our implementation of the CT-HYB QMC solver to analyze these effects.

\subsubsection {Role of spin-flip and pair-hopping interactions}
First we analyze the role of spin-flip and pair hopping interactions
on the orbital melting transition. The minimal Hubbard Hamiltonian which is believed to retain the essential physics \cite{Ahn2000} to study this issue is a two-band Hubbard model for $e_g$ states coupled to disordered $t_{2g}$ spins via
the Coulomb interaction, which acts as a local magnetic field $h=J_{t_{2g}}S_{t_{2g}}$.
Thus in Hamiltonian (\ref{eq:model}) the one-electron term becomes
\begin{equation*}
\left\{
\begin{array}{clc}
\varepsilon_{m \sigma m^\prime\sigma^\prime}&=\left(\varepsilon_{\rm JT}\tau_x^i+\varepsilon_{\rm T} \tau_z^i\right)\delta_{\sigma,\sigma^\prime}-h\sigma_z^i  \\ \\
t^{ii^\prime}_{m\sigma m^\prime \sigma^\prime}&=u_{\sigma,\sigma^\prime} t^{ii^\prime}_{m m^\prime}. \end{array}
\right.
\end{equation*}
The index $m$ runs over the $e_g$ Wannier orbitals $|{x^2-y^2}\rangle$ and $|{3z^2-r^2}\rangle$,
$\sigma_z$ is the Pauli $z$ matrix, while $\tau_x$ and $\tau_z$ are pseudospin operators acting on orbital degrees of freedom ($\tau_z|3z^2-r^2\rangle=1/2 |3z^2-r^2\rangle$, $\tau_z|x^2-y^2\rangle=-1/2|x^2-y^2\rangle$ 
$\tau_x|3z^2-r^2\rangle=|x^2-y^2\rangle$). 
The energies $\varepsilon_{\rm JT}$ and  $\varepsilon_{\rm T}$ yield, repsectively, the Jahn-Teller and tetragonal crystal-field splitting.
Finally $u_{\sigma,\sigma^\prime}=2/3$ is a band renormalization factor which accounts for the disorder in the orientations of the
$t_{2g}$ spins.\cite{Ahn2000} 
For the effective magnetic field $h$, we present calculations for the theoretical estimate\cite{Yamasaki2006}  $h\sim 1.35$~eV;
our results for the orbital-melting transition and the orbital polarization are however
weakly dependent on $h$ in the relevant regime, in which $e_g$ and $t_{2g}$ spins are locally aligned.
For the $e_g$ basis, the Coulomb interaction is composed of
density-density interactions, spin-flip and pair hopping terms.  
We use the theoretical estimates $U_0=5$~eV and $J_{e_g}\sim \tilde{J}\sim 0.76$~eV 
for the $e_g$ screened direct and exchange on-site Coulomb interaction.\cite{Mizokawa1996,Yamasaki2006,Pavarini2010} In order to calculate the critical temperature
due to super-exchange only, we set the crystal-field parameters 
to zero: $\varepsilon_{\rm JT}=\varepsilon_{\rm T}=0$. This disentanglement procedure has been proposed in Ref.~\onlinecite{Pavarini2008}, and was succesfully used
to study orbital order in cuprates and manganites.\cite{Pavarini2008,Pavarini2010,Flesch2012}
\begin{figure}[tb]
\includegraphics[width=0.45\textwidth]{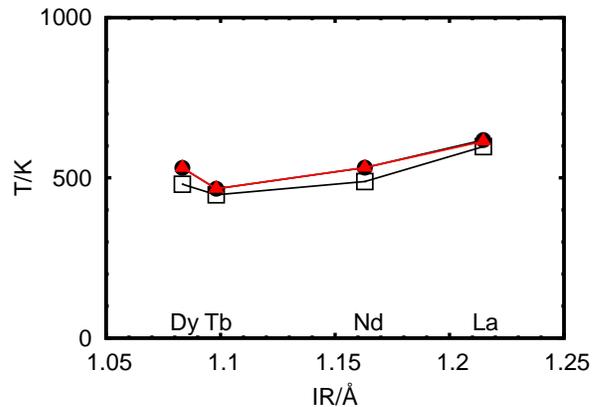}
\caption{\label{fig:tkk}Orbital-order transition temperature due to super-exchange, $T_\textrm{KK}$, versus the $R^{3+}$ ionic radius in the $R\textrm{MnO}_3$ series ($R=\textrm{Dy, Tb, Nd, La}$). Empty symbols: $T_\textrm{KK}$  (total energy gain) taken from
Ref.~\onlinecite{Flesch2012} ; calculations were done for density-density Coulomb interactions and
using a Hirsch-Fye QMC solver.
Filled symbols, grey: CT-QMC (segment solver) and density-density Coulomb interactions only.
Filled symbols, black: CT-QMC (Kryolv solver) and full Coulomb interaction.}
\end{figure}

We show in Fig.\ \ref{fig:tkk} the results of our calculations based on our CT-HYB QMC solver; we use the Krylov approach for the model with spin-flip and pair hopping terms and the segment method for the model with density-density Coulomb terms only. The figure shows the orbital-order transition temperature  due to super-exchange only, $T_\textrm{KK}$, for relevant elements of the series of rare-earth manganites. This figure demonstrates that the spin-flip and
pair hopping terms affect very little the overall trends and even the absolute value of $T_\textrm{KK}$. 
These results all reinforce previous conclusions \cite{Flesch2012} that super-exchange has a small influence in determining the orbital order to disorder transition observed in rare-earth managnites.
\begin{figure}[t]
\rotatebox{270}{\includegraphics[width=0.5\textwidth]{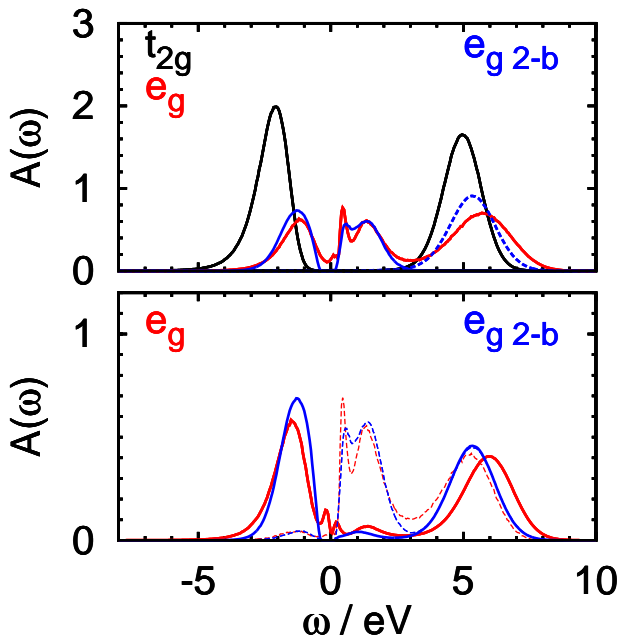}}
\caption{\label{5vs2}  (Color on-line) Top panel: LaMnO$_3$, comparison
of the spectral function matrix obtained for the 5-band Hubbard model
and the $e_g$ 2-band model with classical $t_{2g}$ spins. Calculations have been performed at $T\sim 290$~K and $U_0\sim 5$~eV. The chemical potential $\mu$ is at energy zero for the 2-band model and at $\sim0.3$~eV for the 5-band model.
The Jahn-Teller and tetragonal crystal-field splittings are set to zero.
Black line: $t_{2g}$ spectral function. 
Light lines: $e_{g}$ spectral function from the 5-band model.
Dark lines: $e_{g}$ spectral function from the 2-band model,
spin up (full) and down (dashed).
The position of the spin down Hubbard band depends on the effective magnetic field $h$, i.e. on $J_{t_{2g}}$.
Bottom panel: Comparison of the $e_g$ spectral function matrices, 
orbitally resolved. Full (dashed) lines: most (least) occupied orbital.
Dark (light) lines: 2-band (5-band) model.
}
\end{figure}

\subsubsection{ Classical $t_{2g}$ spins versus full 5-band model for LaMnO$_3$}
Next we test the validity of the classical $t_{2g}$ spin approximation for the orbital-order melting
transition. To do this, we compare  the results of the previous section with those obtained for the full $5$-band Hubbard model described by Hamiltonian (\ref{eq:model}).
To study the orbital order due to superexchange only,
we again set to zero the crystal-field splitting within the $e_g$ doublet and $t_{2g}$ triplet; 
we retain however the cubic crystal field which splits $t_{2g}$ and $e_g$; finally, we 
perform the LDA+DMFT calculations at $T\sim 290$~K, i.e., well below $T_{\rm KK}$. Since we have already shown that spin-flip and pair-hopping do not affect the transition temperature, we neglect them here to speed up the calculations. Furthermore, to compare directly the results of the two- and five-band model, we assume $J_{e_g\mbox{\rm-}t_{2g}} \sim h/S_{t_{2g}}$ for the $e_g$-$t_{2g}$ exchange coupling and neglect other small Coulomb anisotropies. 
The LDA+DMFT calculation for the five-band model yields half-filled  
$t_{2g}$ states and almost fully polarized  $e_g$ states. %
The occupied $e_g$ state $|\theta\rangle=
\cos\frac{\theta}{2}|3z^2-r^2\rangle-
\sin\frac{\theta}{2}|x^2-y^2\rangle
$ is the orbital with $\theta\sim 90^o$,
in excellent agreement with the results from the classical $t_{2g}$ spins approximation,
which gives basically the same state.
The spectral function matrix calculated for the $5$- and $2$-band model are compared in Fig.~\ref{5vs2}. 
This figure shows that not only the orbitals but also, surprisingly, 
the overall spectral function matrices are in good agreement. 
Because the five-band model includes the full dynamic of the $t_{2g}$ electrons,\cite{erik}
the effective $U_0$ is larger than for the two-band model.
By scanning different $U_0$ between 7 eV and 5 eV
we find that $U_0\sim 5.5$~eV yields a gap quite close to that of the two-band model. 
This shows that in the two-band model the Coulomb integral $U_0$ is screened $\sim 10\%$  by the $t_{2g}$ electrons. 
The half-filled $t_{2g}$ bands  exhibit a very large gap because
at half-filling the $t_{2g}$  exchange couplings effectively enhance the effect of the
Coulomb repulsion $U_0$. Finally, we find that the on-site spin-spin correlation
function $\langle S_z^{t_g} S_z^{e_g} \rangle\sim 0.74$, very close to the value 0.75 expected
for aligned $e_g$ and $S_{t_{2g}}=3/2$ $t_{2g}$ spins. For what concerns the sign problem, we find it  negligible for all of these calculations (the average sign is $\sim 0.99$ in the worse case).

\subsection{\label{sec:applications:3band}Orbital fluctuations and magnetism in CaVO$_3$ and YTiO$_3$}
The importance of orbital fluctuations in the physics of $3d^1$ perovskites
has been debated since long.\cite{Khaliulin1,Ulrich,Khaliulin2,Cwick,Pavarini2004,Pavarini2005} Single-site DMFT calculations have shown that in the presence of crystal-field splitting
Coulomb repulsion strongly suppresses orbital fluctuations.\cite{Pavarini2004} However, 
these conclusions were based on a Hubbard model with density-density Coulomb interactions
only. In this section we analyze the effect of the neglected spin-flip and pair-hopping Coulomb interactions. Furthermore, exploiting our efficient CT-HYB solver,
we address the issue of the nature of the low temperature (30 K) \cite{goral,Ulrich}
ferromagnetic transition in YTiO$_3$.

\subsubsection{Orbital fluctuations}
The minimal model to consider for $3d^1$ transition-metal oxides is a three-band
Hubbard model for the $t_{2g}$ bands including  spin-flip and pair hopping terms, and with
\begin{equation*} 
\left\{
\begin{array}{cl}
\varepsilon_{m \sigma m^\prime \sigma^\prime} &=\varepsilon_{m m^\prime} \delta_{ \sigma,\sigma^\prime}
 \\ \\
  t^{ii^\prime}_{m\sigma m^\prime \sigma^\prime}&= t^{ii^\prime}_{m m^\prime}\delta_{ \sigma,\sigma^\prime}
\end{array}
\right.
\end{equation*}
where $m,m^\prime=xy,xz,yz$.
For the Coulomb parameters we use $U_0=5$~eV and $J_{t_{2g}}\sim 0.68$~eV (CaVO$_3$) or $J_{t_{2g}}=0.64$~eV (YTiO$_3$)
from theoretical estimates and previous works.\cite{MF96,Pavarini2004}
Because the local Hamiltonian mixes flavors even in the crystal-field basis, we perform the LDA+DMFT calculations using the Krylov version
of our general CT-HYB QMC solver.
\begin{table}[t]
\begin{ruledtabular}
\begin{tabular}{cccc}
	&	$n_1$	&	$n_2$	&	$n_3$	\\
\hline
CaVO$_3$	&		$0.47$&	$0.28$	&	 $0.25$	\\
YTiO$_3$		&	$0.98$	&	$0.01$	&		$0.01$\\
\end{tabular}
\end{ruledtabular}
\caption{\label{tab:nat_occ_3band}Occupations $n_i$ of the natural orbitals
(with $n_i>n_{i+1}$) at $T=190$~K in CaVO$_3$ and YTiO$_3$ obtained by diagonalizing the occupation matrix.}
\end{table}

In Table \ref{tab:nat_occ_3band} we show the occupations $n_i$ of the natural orbitals at $\sim$190~K in CaVO$_3$ and YTiO$_3$. We find that CaVO$_3$ is a paramagnetic metal with a small orbital polarization. Instead, YTiO$_3$ is a paramagnetic insulator with orbital polarization $p=n_1-(n_2+n_3)/2\sim 1$, i.e. basically full (orbitally ordered state). For this system, the double occupancies at
290~K are small, i.e., we find $\frac{1}{2}\sum_{m\sigma \ne m^\prime \sigma^\prime} \langle \hat{n}_{m\sigma} \hat{n}_{m^\prime\sigma^\prime} \rangle \sim 0.015$ for YTiO$_3$. 
The occupied orbital is 
$0.611|{xy}\rangle-0.056|{xz}\rangle+0.789|{yz}\rangle$. 
We find occupied state
and orbital polarization are  basically the same with full Coulomb and density-density approximation.
Previous  calculations  \cite{Pavarini2004} in which spin-flip and pair-hopping terms
have been neglected and $T\sim770$~K are in line with these results. 
This shows that spin-flip and pair-hopping terms  do not change the conclusion that orbital fluctuations
are strongly suppressed in the Mott insulator YTiO$_3$.
In the CT-HYB QMC simulations the average sign is $\sim 0.9$ for YTiO$_3$ and $\sim 0.95$ for CaVO$_3$.

\subsubsection{Ferromagnetism in YTiO$_3$}
\begin{figure}[tb]
\includegraphics[width=0.5\textwidth]{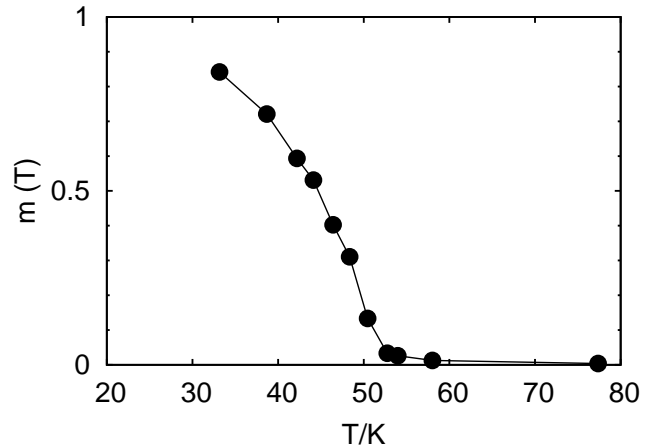}
\caption{\label{mu}
Ferromagnetic spin polarization as a function of temperature in YTiO$_3$.
The plot shows a transition at the critical temperature $T_C\sim 50$~K, slightly overestimating
the experimental value $T_C\sim 30$~K, as one might expect by mean-field calculations.
}
\end{figure}

YTiO$_3$ is one of the few ferromagnetic Mott insulators. 
Neutron scattering experiments pointed out early on the difficulties in
reconciling ferromagnetism and the expected orbital order, \cite{Ulrich} and 
there have been suggestion that the ferromagnetic state could rather be associated
with a quadrupolar order and large scale orbital fluctuations.\cite{Khaliulin2} 
However, second-order perturbation theory calculations indicate
that ferromagnetism and orbital order could be reconciled, provided that the real 
crystal-structure of YTiO$_3$, including the GdFeO$_3$-type distortion (tilting and rotation
of the octahedra, and deformation of the cation cage) is taken into account.\cite{Pavarini2005}  
To clarify this point, we check the instability towards ferromagnetism of the three-band
$t_{2g}$ Hubbard model obtained for the experimental structure of YTiO$_3$. 
With this approach we calculate the ferromagnetic transition
temperature $T_C$ due to super-exchange alone in the orbitally ordered phase. 
Since experimentally $T_C\sim 30$~K, we have to perform LDA+DMFT calculations down to very low temperatures, 
which becomes possible with the CT-HYB QMC solver. 
On lowering the temperature, we find that the sign problem becomes sizable
(average sign $\sim 0.7$ at $40$~K).
However, we can basically eliminate it (average sign $\sim 0.97$) by performing 
the LDA+DMFT calculations in the basis which diagonalize the crystal-field matrix,
even though the hybridization function has off-diagonal terms
of comparable size in the two bases.
In Fig.~\ref{mu} we show the LDA+DMFT magnetization $m(T)$ of the $t_{2g}$ states as a function of the temperature. Remarkably, we find a transition at about 50~K,
in excellent agreement with experiments,\cite{TN} which yield $T_C\sim 30$~K;
the overestimation can be ascribed to the  mean-field approximation, and to the
fact that, since the critical temperature is very small, it is sensitive to tiny details.
The occupied orbital does not change significantly in the magnetic
phase, indicating that the occupied orbital remains the one that diagonalizes the crystal-field 
matrix, i.e., in the magnetic phase there is no sizable change of orbital\cite{Pavarini2010,Flesch2012} due to super-exchange.

\section{\label{sec:conclusions}Conclusions}
We implement an efficient general version of the continuous-time  
hybridization expansion (CT-HYB) quantum Monte Carlo solver,
which allows us to investigate ordering phenomena in
strongly correlated transition-metal oxides in a more
realistic setting.
Our implementation of CT-HYB QMC works for systems of arbitrary symmetry. In cases where symmetry
allows it (i.e., if the local Hamiltonian does not mix flavors)
we use the fast segment solver. In more realistic situations we use the Krylov approach
and, away from phase transition, trace truncation.
We find that in all considered cases the minus sign problem mostly appears when off-diagonal crystal-field terms are present
(and is strongly suppressed in the basis of crystal-field states), while off-diagonal terms of the hybridization function matrix are not as critical.\cite{sign}
We show that spin-flip and pair-hopping terms hardly affect
the strength of the super-exchange orbital-order transition temperature
in rare-earth manganites.
We show that the classical $t_{2g}$ spin approximation
for LaMnO$_3$ works excellently, not only for what concerns
orbital order, but, surprisingly, also for the overall shape of the spectral function
matrix. We show that  spin-flip and pair-hopping terms
also do not change the conclusion that orbital-fluctuations
are strongly suppressed in YTiO$_3$. Furthermore,
we calculate the critical temperature for ferromagnetism
in the orbitally ordered phase, and find excellent agreement with
experiments. This shows that that the predicted 
orbital order is fully compatible with ferromagnetism.
\acknowledgments

Calculations were done on the J\"ulich Blue Gene/Q and Juropa. We acknowledge financial support from the Deutsche Forschungsgemeinschaft through research unit FOR 1346.

\appendix
\section{General CT-HYB solver}
In this appendix we fix the notation and explain the details of our implementation
of the general CT-HYB quantum-impurity solver.
The DMFT quantum-impurity Hamiltonian is  ${H}  =  {H}_\textrm{loc} + {H}_\textrm{bth} + {H}_\textrm{hyb}$, where
\begin{eqnarray*}
H_\textrm{loc}& =& \sum_{\alpha\bar\alpha} 
\tilde
\varepsilon_{\alpha \bar\alpha}
c^{\dagger}_\alpha c^{\phantom{\dagger}}_{\bar\alpha} + \frac{1}{2} \sum_{\alpha\alpha^\prime} \sum_{ {\bar\alpha}{\bar\alpha}^{\prime}} U_{\alpha \alpha^\prime {\bar\alpha}{\bar\alpha}^{\prime}} c^{\dagger}_\alpha c^{\dagger}_{\alpha'}c^{\phantom{\dagger}}_{{\bar\alpha}^{\prime}} 
c^{\phantom{\dagger}}_{{\bar\alpha}}, \\
H_{\rm{bth}} &=& \sum_{\gamma} \epsilon_{\gamma}b^{\dagger}_{\gamma} b^{\phantom{\dagger}}_{\gamma},	\\
H_{\rm{hyb}} &=&\sum_{\gamma}\sum_{\alpha} \left[ V_{\gamma,\alpha } c^{\dagger}_\alpha b^{\phantom{\dagger}}_{\gamma} + h.c. \right].
\end{eqnarray*}
The combined index $\alpha=m \sigma$ labels spin and orbital degrees of freedom (flavors).
For the bath we use, without loss of generality,\cite{CDMFT} the basis 
which diagonalizes ${H}_\textrm{bth}$, with quantum numbers $\gamma$. 
Finally, we define $\tilde\varepsilon_{\alpha \bar\alpha}=\varepsilon_{\alpha \bar\alpha}\! -\!\Delta \varepsilon_{\alpha \bar\alpha}^{\rm DC} $, where $\varepsilon_{\alpha \bar\alpha}$ is the crystal-field
matrix and
$\Delta \varepsilon_{\alpha \bar\alpha}^{\rm DC}$ is the double counting correction;
in the cases considered in the present paper the latter is a shift of the chemical potential $\mu$.

\subsection{Hybridization-function expansion}
By expanding the partition function in powers of ${H}_{\rm hyb}$
and going to the interaction picture 
$H_{\rm hyb}(\tau)={\rm e}^{\tau(H_{\rm bth} +H_{\rm loc})}
 H_{\rm hyb}{\rm e}^{-\tau(H_{\rm bth} +H_{\rm loc})}$ 
with $\beta=1/k_BT$ we obtain the series
\begin{eqnarray*}
Z &=& {\rm Tr} \left[ e^{-\beta ({H}_{\rm bth} +{H}_{\rm loc})}  
{\cal{T}} e^{-\int_0^\beta d\tau H_{\rm hyb}(\tau)}\right]\\
  &=&\sum_{m=0}^\infty (-1)^m %
 \!\! \int^{(m)} \!\!\! \!\!\!\!d{\bm \tau} \;  
{\rm Tr} \; {\cal{T}} \left[ 
{\rm e}^{-\beta ({H}_{\rm bth} +{H}_{\rm loc})} \prod_{i=m}^1
H_{\rm hyb}(\tau_i)\right]
\end{eqnarray*}
where ${\cal {T}}$ is the time order operator, ${\bm \tau}=(\tau_1,\tau_2,\dots \tau_m)$ 
with $\tau_{i+1} \ge \tau_i$ and
\begin{eqnarray*}
\int^{(m)}\!\!\! \!\!\!\!d{\bm\tau}&\equiv& \int_0^\beta d\tau_1 \dots \int_{\tau_{m-1}}^\beta \!\!\! \!\!\! d\tau_m .
\end{eqnarray*}
In the trace only terms containing an equal number of creation and annihilation operators in both the bath and impurity sector, i.e., only even expansion orders $m=2n$ contribute.
Introducing the bath partition function $Z_\textrm{bth} ={\rm Tr}\; {\rm e}^{-\beta{H}_\textrm{bth}}$, the partition function can be factorized
\begin{equation}\label{Z}
\frac{Z} {Z_\textrm{bth}}
 =  
\sum_{n=0}^\infty \int^{(n)}\!\! \!\!\!\!d{\bm \tau}\int^{(n)}\!\! \!\!\!\!\!\!  d{\bar{\bm{\tau}} } \;
\sum_{
      {\bm \alpha} 
  \bar{\bm \alpha} 
      }
      z^{(n)}_{  {\bm \alpha}, \bar{\bm \alpha} } ({\bm \tau},\bar{\bm \tau})\,,
\end{equation}
with 
\begin{eqnarray*}
 z^{(n)}_{{\bm \alpha},\bar{\bm \alpha} } ({\bm \tau},\bar{\bm \tau})
&=&
 t^{(n)}_{{\bm \alpha}, \bar{\bm \alpha} }({\bm \tau},\bar{\bm \tau})\;
 d^{(n)}_{\bar{\bm \alpha}, {\bm \alpha} } ({\bm \tau},\bar{\bm \tau})\,. 
\end{eqnarray*}  
The first factor is the trace over the impurity states
\begin{eqnarray*}
  t^{(n)}_{{\bm \alpha} ,  \bar{\bm \alpha} }({\bm \tau},\bar{\bm \tau}) &=&
 {\rm Tr}\; {\cal{T}} 
 \left[ {\rm e}^{-\beta({H}_\textrm{loc}-\mu N)} 
 \prod_{i=n}^1 c^{\phantom{\dagger}}_{\alpha_i}(\tau_{i}) 
 c^{\dagger}_{\bar\alpha_i}(\bar\tau_{i}) \right], 
\end{eqnarray*}  
where $c_\alpha^{(\dagger)}(\tau)=e^{\tau (H_{\rm loc}-\mu N)} c_\alpha^{(\dagger)} e^{-\tau (H_{\rm loc}-\mu N)}$ and $N$ is the total
number of electrons on the impurity.
For expansion order $m=2n$, the vector ${\bm \alpha}=(\alpha_1,\alpha_2\dots\alpha_n)$ gives the flavors $\alpha_i$ associated with the $n$ annihilation operators on the impurity at imaginary times $\tau_i$, while the
$\bar{\bm \alpha}=(\bar\alpha_1,\bar\alpha_2\dots\bar\alpha_n)$
are associated with the $n$ creation operators at $\bar\tau_i$.
The second factor is the trace over the non-interacting bath, which is given by the determinant
\begin{eqnarray*}  
d^{(n)}_{\bar{\bm \alpha},{\bm \alpha} } 
  ({\bm \tau},\bar{\bm \tau})&=& \det [F^{(n)}_{\bar{\bm \alpha},{\bm \alpha} } ({\bm \tau},\bar{\bm \tau})]
\end{eqnarray*}
of the $n\times n$ square hybridization-function matrix  with matrix elements $[F^{(n)}
_{\bar{\bm \alpha},{\bm \alpha} }({\bm \tau},\bar{\bm \tau})]_{i^\prime,i}     = F_{\bar\alpha_{i^\prime}\alpha_i} (\bar\tau_{i^\prime}-\tau_i)$ given by
\begin{eqnarray*}\label{eq:hyb}
F_{\bar\alpha \alpha}(\tau) &=& \sum_{\gamma} \frac{V_{\gamma,\bar\alpha} \bar{V}_{\gamma,\alpha}}{1+{\rm e}^{-\beta\epsilon_{\gamma}}} \times
\begin{cases}
-{\rm e}^{-\epsilon_{\gamma}\tau}	&	\tau> 0	\\
{\rm e}^{-\epsilon_{\gamma}(\beta+\tau)}	&	\tau < 0	.\nonumber
\end{cases}
\end{eqnarray*}
On the Fermionic Matsubara frequencies, $\omega_n$, its Fourier transform
\begin{eqnarray*}
F_{\bar\alpha \alpha}(\omega_n) = \sum_{\gamma} \frac{V_{\gamma,\bar\alpha } \bar{V}_{\gamma,\alpha}}{i \omega_n - \epsilon_{\gamma}}
\end{eqnarray*}
is related to the bath Green-function matrix $\mathcal{G}$ by 
\begin{eqnarray*}
F_{\bar\alpha\alpha}(\omega_n) = i\omega_n \delta_{\bar\alpha\alpha} \!-\! \tilde\varepsilon_{\bar\alpha\alpha}\! -\! (\mathcal{G})^{-1}_{\bar\alpha\alpha}(\omega_n),
\end{eqnarray*}
as can be shown by downfolding\cite{erikbook2011} 
\begin{eqnarray*}
(\mathcal{G})^{-1}(\omega_n)=
\left( \begin{array}{c|cccc}
i\omega_nI_0 -H_0 	
&	 V_{1,0}	&	V_{2,  0}	&	\hdots	\\ \hline
 \bar V_{1,0}	&	i\omega_n-\epsilon_1	&	0	&	\hdots	\\
 \bar V_{2,0}	&	0	&	i\omega_n-\epsilon_2	&  \hdots	\\
\vdots	&	\vdots	&	\vdots	&	\ddots \\ 

\end{array} \right)
\end{eqnarray*}
to the impurity block ($i=0$). 
Here the matrix elements of $H_0$ and $I_0$ are given by $(H_0)_{\alpha\bar\alpha} =
\tilde
\varepsilon_{\alpha \bar \alpha}$ and $(I_0)_{\alpha\bar\alpha}=\delta_{\alpha,\bar\alpha}$, while 
$\left(V_{0,i}\right)_{\bar \alpha i}=V_{\bar\alpha,i}$, and $\left(\bar V_{i,0}\right)_{i\alpha }=\bar{V}_{i,\alpha}$.

To speed up the calculations, we exploit symmetries.
If $N_b$ blocks of flavors are decoupled by symmetries, the hybridization function matrix
is block-diagonal in those flavors. We then write the partition function in terms of the expansion orders $n_b$ in each block, with $n=\sum_{b=1}^{N_b} n_b$,
${\bm \tau}=\sum_{b=1}^{N_b} {\bm \tau}_b$, and ${\bm \alpha}=\sum_{b=1}^{N_b} {\bm \alpha}_b$.
Thus
\begin{eqnarray}
\nonumber
\frac{Z} {Z_\textrm{bth}} &=& \left[ \prod_{b=1}^{N_b}  \sum_{n_b=0}^\infty 
\int^{(n_b)} \!\!\!\!\!d{\bm \tau}_b \int^{(n_b)}\!\!\!\!\! d\bar{\bm \tau}_b   \;
  \sum_{{\bm\alpha}_b\bar{\bm\alpha}_b}  \; \right]  
  z^{(n)}_{\bm\alpha,\bar{\bm \alpha} }({\bm \tau},\bar{\bm \tau}) 
 \end{eqnarray}
with
 \begin{eqnarray}
 \nonumber
d^{(n)}_{ \bar{\bm \alpha}, 
  {\bm \alpha}  }({\bm \tau},\bar{\bm \tau})&=& 
\prod_{b=1}^{N_b} 
 d^{(n_b)}_{\bar{\bm \alpha}_b,  
  {\bm \alpha}_b} ({\bm \tau}_b,\bar{\bm \tau}_b)
\end{eqnarray}
and
 \begin{eqnarray*} 
  t^{(n)}_{{\bm \alpha}, 
  \bar{\bm \alpha}} ({\bm \tau},\bar{\bm \tau})&=&
 {\rm Tr} \; {\cal{T}} 
 \left[ {\rm e}^{-\beta({H}_\textrm{loc}-\mu N)}  \prod_{b={1}}^{N_b}
 \prod_{i={n_b}}^1 c^{\phantom{\dagger}}_{\alpha_{bi}}(\tau_{bi}) 
 c^{\dagger}_{\bar\alpha_{bi}}(\bar\tau_{bi}) \right].
\end{eqnarray*}

\subsection {Segment solver and Krylov approach}
Calculating the trace over the impurity states involves propagating states in the impurity Hilbert space. For models with many orbitals this can become very demanding. We therefore
use a multi-approach scheme. %
When the on-site Hamiltonian conserves the flavors we use the so-called segment approach,\cite{Werner2006} which is extremely fast. 
In such cases only terms with an equal number of creation and annhilation 
operators {\em per flavor} contribute to the local trace, and it is convenient to express the 
partition function in expansion orders $n_a$ for flavors $a$. %
The partition function then can be rewritten as
\begin{eqnarray}
\label{zfunctionnew}\nonumber
\frac{Z} {Z_\textrm{bth}} &=&
\left[\prod_{a=1}^{N_a}\sum_{n_a=0}^\infty 
\int^{(n_a)} \!\!\!\!\! d{\bm \tau}_{a} 
\int^{(n_a)} \!\!\!\!\! d\bar{\bm \tau}_{a} \right] z^{(n)}_{ \bm\alpha,\bar{\bm \alpha}  }({\bm \tau},\bar{\bm \tau}). 
\end{eqnarray}
Here 
$    {\bm \tau}=\sum_{a=1}^{N_a} {\bm \tau}_a$  and
     $\bar{\bm\tau} =\sum_{a=1}^{N_a} \bar{\bm \tau}_a$, while the vectors
$    {\bm \alpha}=\sum_{a=1}^{N_a} {\bm \alpha}_a$  and
     $\bar{\bm\alpha} =\sum_{a=1}^{N_a} \bar{\bm \alpha}_a$ have the 
     $n_a$ components $\alpha_{ai}=\bar{\alpha}_{ai}={a}$.
The local trace factors into
\begin{eqnarray*}
t^{(n)}_{\bm\alpha,\bar{\bm \alpha}}({\bm \tau},\bar{\bm \tau}) &=&
 {\rm Tr} \; {\cal{T}} 
 \left[ {\rm e}^{-\beta({H}_\textrm{loc}-\mu N)}  \prod_{a=1}^{N_a}
 \prod_{i={n_a}}^1 c^{\phantom{\dagger}}_{a}(\tau_{a i}) 
 c^{\dagger}_{a}(\bar\tau_{a i}) \right] \\
 &=&
 \left( \prod_{a=1}^{N_a} s_a^{n_a} \right) e^{-\sum_{a  a^\prime}\left[ 
   (\tilde\varepsilon_{a a}-\mu) \delta_{a, a^\prime}
   +\frac{1}{2} 
 \tilde{u}_{a  a^\prime}\right] l_{ a  a^\prime}},
 \end{eqnarray*}
where 
$l_{a a^\prime}$ is the length of the overlap of the $\tau$ segments $a$ and 
$ a^\prime$, $s_a={\rm sgn} (\tau_{a 1} -\bar\tau_{ a 1})$ is the Fermionic sign,
and 
$\tilde{u}_{a  a^\prime}=U_{a  a^\prime  a^\prime a}+ U_{a a^\prime a  a^\prime }$ 
is the interaction.
\begin{figure}[tb]
\includegraphics[width=0.45\textwidth]{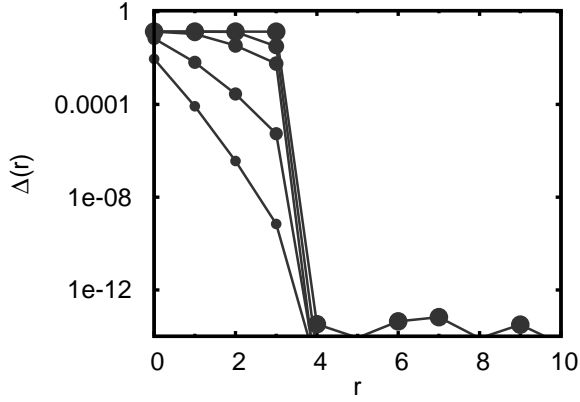}
\caption{\label{fig:krylov} Convergence of the Krylov approximation 
$|\psi(\tau)\rangle_r$ to  $|\psi(\tau)\rangle={\rm e}^{-({H}_{\rm loc}-E_0)\tau}|{\psi}\rangle$
for a representative test case (5-orbital model, half filling). The figure shows the difference $\Delta (r)=|| \psi(\tau)\rangle_r-|\psi(\tau)\rangle|$. 
Symbols (in order of increasing size): $\tau=0.005$, $0.05$, $0.5$, $5$ and $100$. 
}
\end{figure}

In all the cases in which the local Hamiltonian mixes flavors, we adopt the Krylov method.\cite{Lauchli2009} 
At the beginning of the DMFT loop we calculate  all the eigenstates of $H_{\rm loc}$,  $\{ |\Psi_n\rangle \}$, and their energies $\{ E_n \}$; a given state $|\Psi_n\rangle$ is then propagated
with $e^{-\tau_1 E_n}$; the first creation or annihilation operator met
generates a new state $|\Psi\rangle$, which we propagate with $e^{-(\tau_2-\tau_1) H_{\rm loc}}$ obtaining $|\Psi(\tau_2-\tau_1)\rangle$; we repeat the procedure till the last creation or annihilation operator is met. At the core of the procedure are the matrix-vector multiplications and the propagation of vectors.
For the first aspect, we work in the occupation number basis, in which  $H_{\rm loc}$, and the creation and annihilation operators are sparse matrices. Additionally, we arrange the states according to the symmetries\cite{Haule2007,Gull2011} of $H_{\rm loc}$, so that we have sparse block-diagonal matrices and can exploit to the maximum efficient sparse-matrix multiplication algorithms.
We find that this
typically reduces the CPU time by, e.g., about 15\% for a three-band model.
We use the Krylov approach to calculate 
$|{\Psi(\tau)}\rangle={\rm e}^{-{H}_{\rm loc}\tau}|{\Psi}\rangle$.
First we construct the Krylov space of order $r$, $\mathcal{K}_r(|{\Psi}\rangle)$, i.e., the space spanned by  
$|{\Psi}\rangle, {H}_{\rm loc} |{\Psi}\rangle, {H}^2_{\rm loc} |{\Psi}\rangle \dots {H}_{\rm loc}^r |{\Psi}\rangle$. By means of the Lanczos\cite{erikbook2011} technique  we construct
an orthonormal basis for $\mathcal{K}_r(|{\Psi}\rangle)$, $\{|k\rangle \}$; in this basis ${H}_{\rm loc}$ is tridiagonal with eigenstates $\{|l\rangle\}$ and energies $\{\varepsilon_l\}$.
The matrix exponential ${\rm e}^{-{H}_{\rm loc}\tau}$ is approximated by its projection onto the Krylov space, 
$ {\rm e}^{-{H}_{\rm loc}\tau}|\Psi\rangle\sim 
|\Psi(\tau)\rangle_r=
\sum_{l=0}^r e^{-\tau \varepsilon_l} %
|l\rangle
\langle l|\Psi\rangle $. 
This procedure converges very rapidly with $r$, typically for $r$ much smaller than the dimension of
the Hilbert space,\cite{Prelovsek,Hochbruch} as illustrated in Fig.~\ref{fig:krylov}. We find that the convergence slightly deteriorates with increasing $\tau$ and the complexity of the Hamiltonian (realistic Coulomb vertex, crystal-field matrix), but typically  $2$-$3$ steps are sufficient to obtain accurate results.  
To best exploit the power of the method, we keep $r$ flexible.
Furthermore, to avoid that the norm of the state becomes very large during the propagation, we set $E_0$ to zero,
i.e., substitute $e^{-\tau H_{\rm loc}}$ with $e^{-\tau (H_{\rm loc}-E_0)}$.
In addition, the procedure (propagation and creation/annihilation) is carried out from both the left and the right side of the trace, to minimize
the work needed to measure, e.g., the Green function matrix.
Finally, at low temperatures or far from phase transitions we use the eigenvalues of $H_{\rm loc}$  to determine the relevant energy window and truncate adaptively the outer bracket of the trace. 
This further reduces the CPU time.

\begin{figure}[tb]
\includegraphics[width=0.5\textwidth]{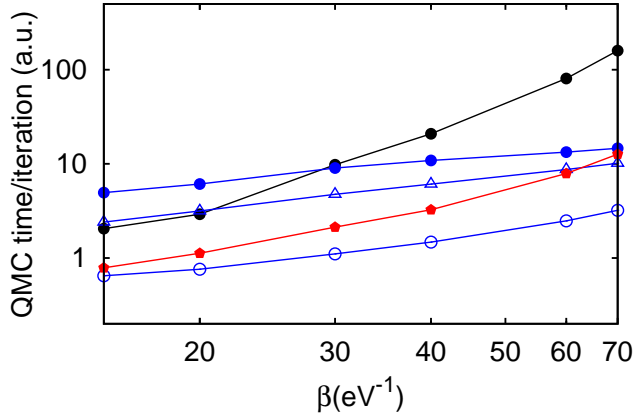}
\caption{\label{scaling} 
(Color on-line) Scaling of our CT-HYB QMC LDA+DMFT code on BlueGene/Q.
Black line: Hirsch-Fye Quantum Monte Carlo solver,
2 orbitals.  
Other lines: CT-HYB Krylov (dark) and CT-HYB segment (light).
Symbols: Two- (circles), three- (triangles) and five- (pentagons) band model. 
Open symbols: Truncated local trace.
All points correspond to calculations of high quality (and with comparable error bars)
for the systems considered in this work.
}%
\end{figure}

The performance of our CT-HYB QMC solver (Krylov and segment version) on the J\"ulich BlueGene/Q, and comparison with Hirsch-Fye QMC, is shown in Fig.~\ref{scaling}.

\subsection{Green-function and occupation matrix}
The partition function (\ref{Z}) can be seen as the sum over all configurations $c=\{\alpha_i \tau_i, 
\bar\alpha_i \bar\tau_i, n\}$ 
in imaginary time and flavors. In a compact form
\begin{eqnarray*}
Z %
&=&\sum_c \left\langle Z %
\right\rangle_c=\sum_c w_c 
\sim\sum_{ \{ c\} } {\rm sign}(w_c)\,,
\end{eqnarray*}
where in the last term the sum is over a sequence of configurations $\{c\}$ sampled by Monte Carlo using $|w_c|$ as the probability of configuration $c$. 
In the segment solver approach, we parametrize the configurations by intervals
$[0,\beta)$ (time-line), occupied by a sequence of creators and annihilators, which define segments on the time-line.
The basic Monte Carlo updates are addition and removal of segments, antisegments or full lines.\cite{Werner2006} 
In the Krylov solver approach  we use  the insertion and removal of pairs of creation and annihilation operators\cite{Werner2006a,Haule2007} as basic updates. In addition, we shift operators in time,\cite{Werner2006,Haule2007} and exchange the configurations of blocks or flavors \cite{Poteryaev2008} 
(global moves).  Finally, a generic observable $O$ can then be obtained as Monte Carlo average
\begin{eqnarray*}
O&\sim&\frac{   
\sum_{ \{ c \}} \langle O\rangle_c \; {\rm sign} (w_c)} 
{\sum_{\{ c \}} {\rm sign}(w_c) }
\end{eqnarray*}
where $\langle O \rangle_c$ is the value of the observable for configuration $c$, and $c$ runs over the configurations visited 
with probability $|w_c|$ during the sampling.
The average expansion order increases linearly with the inverse temperature. For the case of YTiO$_3$, at $\sim 40$~K, the average expansion order is $n\sim40$. 

We calculate the Green function matrix in two ways, directly\cite{Werner2006,Gull2011}  and via Legendre polynomials.\cite{Boehnke2011} 
In the first approach, the Green function matrix is obtained
as Monte Carlo average with $\langle O \rangle_c = \langle G_{\alpha\bar\alpha} \rangle_c $, and
\begin{eqnarray*}
\langle G_{\alpha\bar\alpha} \rangle_c  &=&\sum_{b=1}^{N_b} \sum_{i,j=1}^{n_b} 
 \Delta(\tau,\tau_{bj}\!-\!\bar\tau_{bi}) [M^{(n_b)}]_{bj,bi} \delta_{\alpha_{bj} \alpha} \delta_{\bar\alpha_{bi} \bar\alpha}.
\end{eqnarray*}
Here $M^{(n)}=[F^{(n)}]^{-1}$ is the inverse of the hybridization-function matrix,
which we update at each accepted move, while $\Delta$ is given by
\begin{eqnarray*}
\Delta(\tau,\tau^\prime) &=&-\frac{1}{\beta}
\begin{cases}\phantom{-}
\delta\left(\tau-\tau^\prime\right)	&	\tau^\prime > 0	\\
-\delta\left(\tau-(\tau^\prime+\beta)\right)	&	\tau^\prime < 0
\end{cases}
\end{eqnarray*}
and the $\delta$-function is discretized.
In the second approach, we calculate the Legendre coefficients $\langle O \rangle_c = \langle G_{\alpha\bar\alpha}^l\rangle_c$, with
\begin{eqnarray*}
 \langle G_{\alpha\bar\alpha}^l\rangle_c &=& \sum_{b=1}^{N_b} \sum_{i,j=1}^{n_b} 
{P}_l(\tau_{bj}-\bar\tau_{bi})[M^{(n_b)}]_{bj,bi} \delta_{\alpha_{bj} \alpha} \delta_{\bar\alpha_{bi} \bar\alpha} \\
P_l(\tau)&=&-\frac{\sqrt{2l+1}}{\beta} \times \left\{
\begin{array}{rr} p_l(x(\tau)), & \tau>0\\
           -p_l(x(\tau+\beta)), &\tau<0
\end{array}\right.
\end{eqnarray*}
where $p_l(x)$ is a Legendre polynomial of rank $l$, with $x(\tau)=2\tau/\beta-1$, and we reconstruct the Green function matrix from
\begin{eqnarray*}
 G_{\alpha\bar\alpha}(\tau) &=& \sum_{l=0}^{\infty} \frac{\sqrt{2l+1}}{\beta} p_l(x(\tau))G_{\alpha\bar\alpha}^l. 
\end{eqnarray*}
For what concerns occupations, in the segment solver we calculate them from the total length of the segments of the different flavors; \cite{Werner2006} in the Krylov solver we obtain them 
in two ways, directly from the Green's function and
by explicitly inserting the occupation number operator at the center  of the operator sequence ($\tau=\beta/2$)and calculating the corresponding trace.\cite{Werner2006a,Lauchli2009}
The off-diagonal elements of the local occupation matrix $\langle {c^{\dagger}_{\alpha}c^{\phantom{\dagger}}_{\bar\alpha}}\rangle$, which cannot be obtained by inserting the corresponding operators at $\tau=\beta/2$,\cite{ergodic}
are extracted from the Green function matrix only. 


\begin{thebibliography}{9999}
\bibitem{sign} For the cases in which the hybridization function is diagonal in the flavors,
and the local Hamiltonian does not mix flavors, there is no sign problem.
To see this, we observe that 
%
the sign of the local trace is $\prod_{\alpha=1}^{N_a} s_a^{n_a}$, as shown
in the Appendix, section 2. 
The determinant of the hybridization-function matrix factorises in the
product  $ d^{(n)}_{ \bar{\bm \alpha}, 
  {\bm \alpha}  }({\bm \tau},\bar{\bm \tau})= 
\prod_{a=1}^{N_a} 
 d^{(n_a)}_{\bar{\bm \alpha}_a,  
  {\bm \alpha}_a} ({\bm \tau}_a,\bar{\bm \tau}_a)
$.
We then factor out the Fermionic sign $s_a^{n_a}$ of each
$d^{(n_a)}_{\bar{\bm \alpha}_a,  
  {\bm \alpha}_a} ({\bm \tau}_a,\bar{\bm \tau}_a)$; the remaining term $s_a^{n_a}d^{(n_a)}_{\bar{\bm \alpha}_a,  
  {\bm \alpha}_a} ({\bm \tau}_a,\bar{\bm \tau}_a)$ is positive, as can be shown by transforming the bath Hamiltonian for flavor $a$ into a one-dimensional chain, as done in Ref.~\onlinecite{sign2} for the Anderson model and the Hirsch-Fye algorithm.
\bibitem{ergodic} 
{For observables that cannot be expressed as expectation values of density operators,
measuring by insertion at $\beta/2$  is not necessarily ergodic. This is in particular apparent for, e.g., $\langle c^\dagger_{\alpha}(\tau) c_{\bar\alpha}(\tau)\rangle$ with $\alpha\neq\bar\alpha$; if the 
off-diagonal crystal-field matrix elements are zero, by inserting  $c^\dagger_{\alpha}(\beta/2) c_{\bar\alpha}(\beta/2)$ in the local trace we obtain $\langle c^\dagger_{\alpha}(\tau) c_{\bar\alpha}(\tau)\rangle=0$, even if the hybrization-function matrix is not
diagonal.
}
\end{thebibliography}
\end{document}